\documentclass[12pt,epsf]{amsart}
\usepackage{amssymb,amstext}
\usepackage{epsfig}
\newtheorem{theorem}{Theorem}[section]
\newtheorem{lemma}[theorem]{Lemma}

\theoremstyle{definition}
\newtheorem{definition}[theorem]{Definition}

\numberwithin{equation}{section}
  
  \newcommand{\RR}{\mathbb R}
  
  \newcommand{\CC}{\mathbb C}
  \newcommand{\ZZ}{\mathbb Z}
  \newcommand{\TT}{\mathbb T}

  \newcommand{\bs}{\bigskip}
  
  \newcommand{\eb}{\hfill q.e.d.}

 \newcommand{\bew}{{\em Proof:}}

 \newcommand{\M}{\mathcal M}
 
 \newcommand{\F}{\mathcal F}
\newcommand{\D}{\mathcal D}

 \renewcommand{\P}{\mathcal P}
 \newcommand{\PT}{\mathcal T_P}
\newcommand{\im}{\mbox{\rm im\,}}

\newcommand{\spO}{\check\Omega_{\P}}

\newcommand{\spH}{\check H_{\P}}
\newcommand{\cH}{\check H}
\newcommand{\ch}{\v{C}}
\newcommand{\flc}{finite local complexity}

\begin{document}

\title{Pattern equivariant functions and cohomology}

\author{Johannes Kellendonk}
\address{School of Mathematics, Cardiff University, Cardiff CF2 4YH, Wales}

\email{kellendonkj@cf.ac.uk}

\date{\today}


\begin{abstract}
The cohomology of a tiling or a point pattern has originally been
defined via the construction of the hull or the groupoid associated
with the tiling or the pattern. Here we present a construction which is more
direct and therefore easier accessible. It is based on generalizing 
the notion of equivariance from lattices to point patterns of finite
local complexity. 
\end{abstract}

\maketitle

\section{Introduction}

When topological invariants of tilings were first defined and computed
\cite{Co, Cone, Ke2, Ke5, AP} they arose as $K$-groups of 
associated $C^*$-algebras but it became soon clear that, apart from
the order on $K_0$, the $K$-groups are isomorphic to the integer valued 
\ch ech-cohomology of the continuous hull of the tiling, or
equivalently, to the integer valued cohomology of the discrete tiling
groupoid \cite{FoHu}. Neither the continuous hull nor the discrete groupoid
of a tiling are mathematical objects which are easily accessible by
the non-expert. The purpose of this note is to
present a formulation of (real valued) tiling or
point pattern cohomology which we believe is easier to understand on
an intuitive level,
because it involves more standard mathematical objects.
This does not mean that it is easier to compute the
tiling cohomology in this formulation. 
But we hope that it helps
understanding what the cohomology of a tiling actually
means. 

We formulate our results below for Delone sets of
finite local complexity which we call for short here point patterns. This
covers then the case of tilings of finite local complexity since
the topological invariants mentionned depend only on MLD-classes and
tilings are mutually locally derivable from Delone sets.

We will proceed along the following lines: 
Let $\P$ be a tiling or a
point pattern in the Euclidean space $\RR^n$. 
We construct the (real valued) cohomology of $\P$ as the
cohomology of the sub-complex of the de-Rham complex over $\RR^n$
given by the $\P$-equivariant forms. The main new ingredient is thus
the notion of $\P$-equivariance.

\section{$\P$-equivariant functions and cohomology}

In a frequently used approach to describe particle motion in
aperiodic media one investigates a Hamiltonian of the form
$ H = -\frac{\hbar^2}{2m}\Delta + V$
on $L^2(\RR^n)$ where $\Delta$ is the Laplacian (or even the magnetic Laplacian
in the presence of an external magnetic field) and $V$ is a potential
which describes the local interaction between the particle and
the medium. This potential incorporates the aperiodic structure of the material
in such a way that $V(x)$ depends only on the local
configuration one finds around $x$. So if $\P$ is a point pattern or a
tiling representing the aperiodic structure (e.g.\
$\P$ is the set of equilibrium positions of the atoms in the material)
then $V$ should be a $\P$-equivariant function in the following sense.

Recall that a discrete point set has finite local complexity
if up to translation there are only finitely many $r$-patches for any $r$,
an $r$-patch being $B_r(x)\cap \P$,
the intersection of an $r$-ball at some point $x\in\RR^n$ with $\P$.
This implies that $\P$ is uniformly discrete, i.e.\ that
any two distinct points have distance larger than a given $r>0$.
A Delone set is a uniformly discrete point set for which relatively
dense, i.e.\ there 
exists an $R>0$ such all $R$-patches contain at least one point.
\begin{definition}\label{def1}
Let $\P$ be a subset of $\RR^n$ of \flc. We call a function
$f:\RR^n\to X$ into some set $X$
strongly $\P$-equivariant if there exists an $r>0$ such that
$$B_r\cap (\P-x)= B_r\cap (\P-y)\quad \mbox{implies} \quad f(x)=f(y).$$
($B_r$ is the closed $r$-ball around $0$). 
\end{definition}
Suppose that $\D$ is locally derivable from $\P$ in the sense of
\cite{BSJ}. This
means that for all $r>0$ exists $R>0$ such that 
$B_R\cap (\D-x)= B_R\cap (\D-y)$ implies $B_r\cap (\P-x)= B_r\cap (\P-y)$. 
It is immediate from this definition that then any strongly
$\D$-equivariant function is also strongly $\P$-equivariant. In particular
the concept of $\P$-equivariance is defined not only for a single point
set but for MLD classes of point sets.

Recall that the de-Rham complex over $\RR^n$ is the complex
$$
\Omega^0(\RR^n)\stackrel{d}{\to}\Omega^1(\RR^n)\stackrel{d}{\to}\cdots 
\stackrel{d}{\to}\Omega^n(\RR^n)
$$
where $\Omega^k(\RR^n)$ are the differential $k$-forms over $\RR^n$
and $d$ is the exterior derivative. Using standard coordinates 
on $\RR^n$ a differential $k$-form can be written as
$\sum_{i_1,\dots,i_k} f_{i_1\dots i_k}
dx_{i_1}\cdots dx_{i_k}$ with smooth functions
$ f_{i_1\dots i_k}:\RR^n\to\RR$. 
Setting $X=\RR$ we may therefore consider  
strongly $\P$-equivariant differential forms over $\RR^n$ as those for
which the functions $ f_{i_1\dots i_k}$ are smooth and
strongly $\P$-equivariant. We denote them by $\spO(\RR^n)$. Clearly
$d(\spO(\RR^n))\subset \spO(\RR^n)$ and so 
$$
\spO^0(\RR^n)\stackrel{d}{\to}\spO^1(\RR^n)\stackrel{d}{\to}\cdots 
\stackrel{d}{\to}\spO^n(\RR^n)
$$
is a differential sub-complex of the de-Rham complex.
\begin{definition}
The $\P$-equivariant de Rham cohomology of $\RR^n$ is the cohomology 
of the sub-complex defined by  $\spO(\RR^n)$. We denote it by
$\spH(\RR^n)$, i.e.\  
$$\spH^k(\RR^n)=(\ker d\cap \spO^k(\RR^n)) / (\im
d\cap \spO^k(\RR^n)).$$ 
\end{definition}
The following theorem
connects the above definition with the cohomology of $\P$ as
defined as \ch ech cohomology of the
continuous hull of $\P$. 
Its proof will be given elsewhere
\cite{KPneu}
although we give some explanation in Section~\ref{sect4}. 
\begin{theorem}\label{thm1} Let $\P\subset \RR^n$ be a Delone set of \flc.
The real valued \ch ech cohomology of the continuous hull of
$\P$ is isomorphic to the
$\P$-equivariant de Rham cohomology of $\RR^n$.
\end{theorem}

\section{Examples}
Let us present some examples of $\P$-equivariant cohomology
classes  for $\P=\PT$, a Penrose tiling, which may for our purposes be
identified with the point set given by the centers of mass of its tiles. 
In the triangle version $\PT$ 
has $40$ different translational congruence classes of
tiles which are the decorated triangles of Figure~7 and their images under
$10$-fold rotation. Together with a reflection, e.g.\ the reflection
in the $x$-axis, the $10$-fold rotations form symmetry group which
is the dihedral group $D_{10}$.
We call the translational congruence classes of
tiles prototiles and denote
by $t_{10 k+l}$ the triangle corresponding the one in Figure~7.k
rotated about $\frac{l\pi}{5}$ left around.

\epsffile[0 0 430 100]{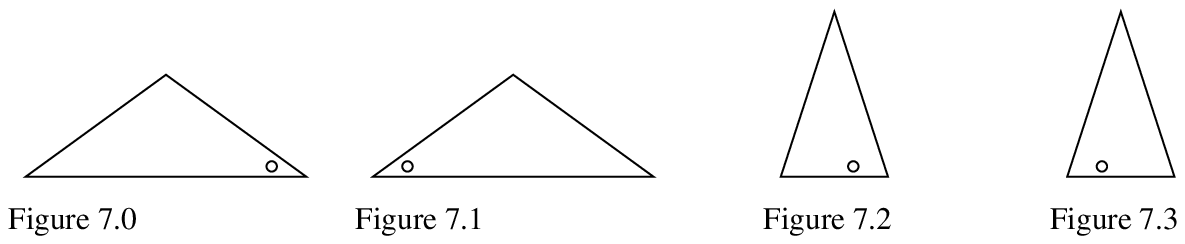}

Since the tiling is two-dimensional 
$\cH_{\PT}^k(\RR^2)$ is non-trivial only in $k=0,1,2$. The case $k=0$
is simple, in fact for any point pattern in any dimension holds 
$\cH_\P^0(\RR^n)=\ker d\cap \spO^0(\RR^n) = \RR$.

\subsection{$k=2$}
A closed strongly $\PT$-equivariant $2$-form is
given by $f dx_1dx_2$ where $f$ is a smooth strongly $\PT$-equivariant function.
Such functions can be constructed as follows: Take a prototile
$t_i$ and let $x_i$ be its center of mass.
Let $\delta_{t_i}$ be the Dirac comb
associated with $t_i$ by which we mean the sum of $\delta$-functions at
points $y\in \RR^n$ such that $t_i-x_i$ occurs at $\PT-y$. 
In other words $\delta_{t_i}$ is a Dirac comb supported on the center
of masses of all tiles which are
translationally congruent to prototile $t_i$.
Furthermore,
let $\rho:\RR^2\to \RR$ be a smooth function of compact support. Then
the convolution product
$f=\delta_{t_i}*\rho$ is a smooth strongly $\PT$-equivariant function.
In fact, any smooth strongly $\PT$-equivariant function can be
approximated in such a way if one allows for Dirac combs which are
supported on arbitrary points sets which are locally derivable from
$\PT$.
The difficult part is to determine when two
$2$-forms differ by a strongly $\PT$-equivariant exact form. It follows
from the Poincar\'e Lemma for compactly supported de Rham cohomology
that different choices of $\rho$ yield the same cohomology class
as long as the average hight $\overline{\rho}:=\int_{\RR^2} \rho dx_1dx_2$ is
kept fixed. We will therefore define 
$\alpha_i(\overline{\rho})=\delta_{t_i}*\rho dx_1dx_2$ 
remaining ambigious about the precise form of $\rho$.  

To go further we have to combine Theorem~\ref{thm1} with known results.
Recall that the Penrose tiling is a substitution tiling and that its
substitution can be used to compute its second cohomology as a
quotient of the cohomology of the AF-groupoid $\mathcal R_\Sigma$
defined by the substitution in \cite{Ke5}\footnote{The highest non-vanishing
cohomology group is
called the group of coinvariants in \cite{Ke5}.}.
More precisely, the so-called substitution matrix defines an endomorphism
$\sigma:\ZZ^{40}\to \ZZ^{40}$ and one finds the
integer second cohomology of the Penrose tiling as a quotient
$\ZZ^{40}/\Gamma$. The real second cohomology of
the Penrose tiling, which we denote here by $H^2(\PT,\RR)$,
 can therefore be identified with a complement of
the real span $\RR \Gamma$ of the sub-group $\Gamma\subset\ZZ^{40}$ 
in $\RR^{40}$. Such a complement 
has been computed \cite{Ke5} 
$$H^2(\PT,\RR)\cong \RR^8 = E(\tau_+^2)\oplus E(\tau_-^2)\oplus
E(-\tau_+)\oplus E(-\tau_-)$$ 
where 
$$\tau_\pm = \xi_\pm+\xi_\pm^{-1} =
\frac{1\pm\sqrt{5}}{2},\qquad\xi_+=e^{\frac{\pi
    i}{5}},\quad \xi_-=e^{\frac{3\pi i}{5}}$$
and $E(s)$ denotes the eigenspace of $\sigma$ (extended to a linear
map on $\RR^{40}$) corresponding to the eigenvalue $s$. 

We can use this to present specific strongly $\PT$-equivariant
$2$-forms which form a linear base for $\cH_{\PT}(\RR^2)$.
In order to do so we
first note that we can identify the standard base $\{e_i\}_i$ of
$\ZZ^{40}$ with indicator functions on certain cylinder sets of
the discrete hull. These cylinder sets are in one-to-one
correspondence with the prototiles and Theorem~\ref{thm1} allows us to
identify the class of $e_i$ in $\RR^{40}/\RR \Gamma$ with the cohomology
class of the two-form $\alpha_i(1)$. All we have to do is therefore to
determine a base for the above complement and this is best done by a
symmetry analysis similar to that given in \cite{OrmesRadinSadun02}.

In terms of the $10$-fold rotation matrix $\omega$
the substitution matrix $\sigma$ is given by
$$
\sigma  =  \left(
\begin{array}{cccc}
\omega^4 & \omega^0 & 0 & \omega^6 \\
\omega^0 & \omega^6 & \omega^4 & 0 \\
\omega^3 & 0 & \omega^7 & 0 \\
0 & \omega^7 & 0 & \omega^3 
\end{array}
\right). 
$$
The eigenspaces $E(\tau_\pm^2)$ are one dimensional and
spanned by 
\begin{equation}\label{eq1}
(\tau_\pm^2{\bf 1},\tau_\pm^2{\bf 1},{\bf 1},{\bf 1})^T
\end{equation}
where ${\bf 1}=(1,\cdots,1)$, an eigenvector of $\omega$ to eigenvalue
$1$. Both eigenvectors are invariant under the
$D_{10}$-symmetry. 
The spaces $E(-\tau_\pm)$ are $3$-dimensional. 
Each decomposes into a $1$-dimensional and a $2$-dimensional real irreducible
representation of $D_{10}$. 
The $1$-dimensional
irreducible real $D_{10}$-module is spanned by
\begin{equation}\label{eq2}
((1-\tau_\pm){\bf a},(\tau_\pm-1){\bf a},-{\bf a},{\bf a})^T
\end{equation} 
where ${\bf a}=(1,-1,\cdots,1)$ is an eigenvector of $\omega$ to eigenvalue
$-1$. Using the matrix $S$ which implements the reflection in the $x$-axis
(see \cite{Ke5}) one sees that the above two vectors are both
invariant under it.
The remaining $2$-dimensional parts of $E(-\tau_\pm)$ are easiest
described if 
we complexify these spaces to observe that each splits into
two $1$-dimensional complex irreducible $D_{10}$-modules, one spanned by
\begin{equation}\label{eq3}
b_\pm=(\tau_\pm\xi_\pm^2{\bf z_\pm},\tau_\pm\xi_\pm^{-1}{\bf z_\pm},
\xi_\pm{\bf z_\pm},{\bf z_\pm})^T 
\end{equation} 
and the other by the complex conjugate $\bar b_\pm$. 
Here ${\bf z_\pm}=(1,{\xi_\pm},{\xi_\pm^2},\cdots,{\xi_\pm^9})$, 
an eigenvector of $\omega$ to eigenvalue
$\xi_\pm^{-1}$ and the reflection
$S$ acts as $Sb_\pm=-\xi_\pm\bar b_\pm$.
We illustrate this result in Figure~2 
by marking the coefficients of $b_\pm$ and $\bar b_\pm$ into
a figure of all prototiles using their identification with the base of
$\ZZ^{40}\subset\CC^{40}$. 

\epsffile[0 0 430 450]{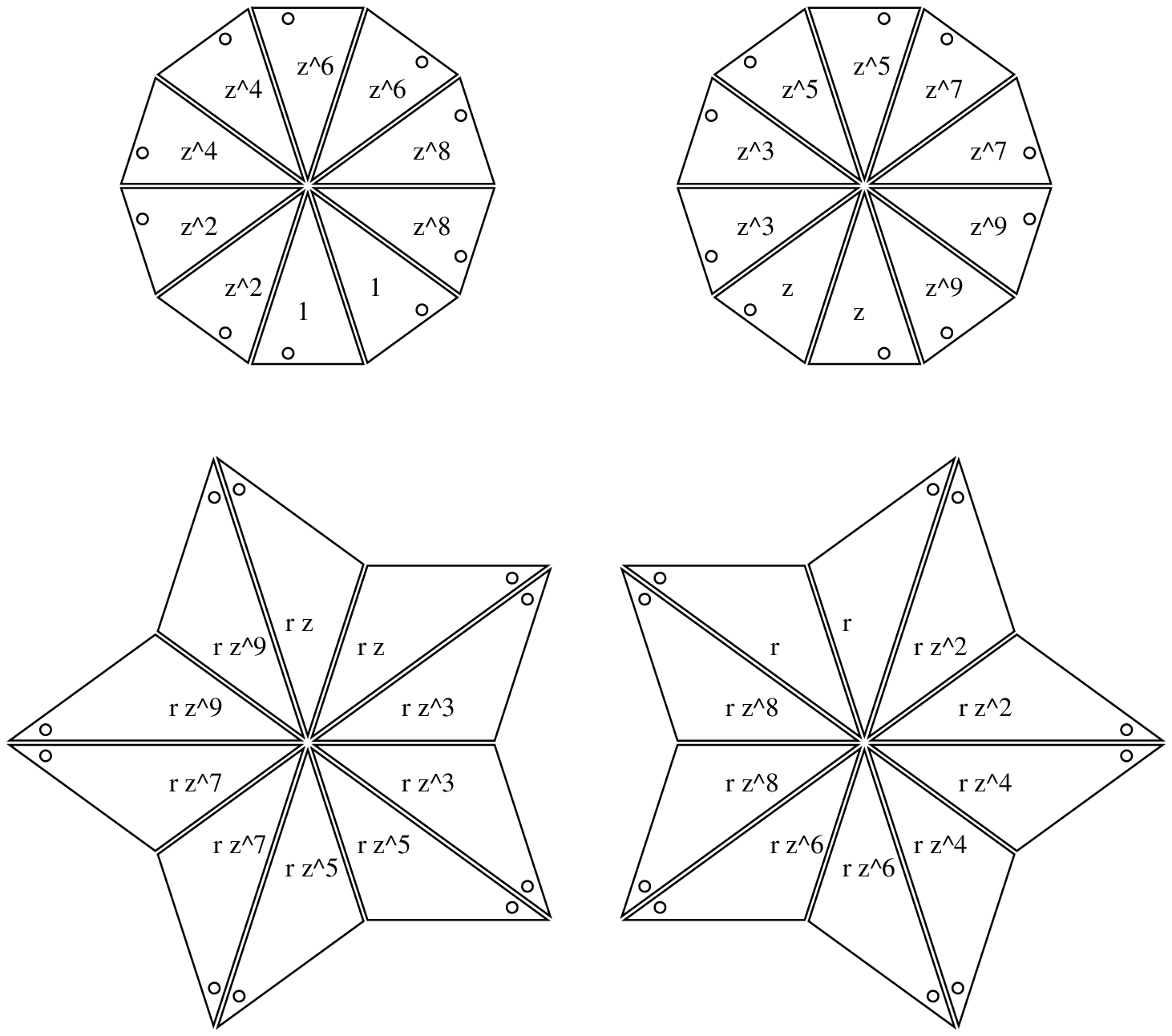}

\noindent
Figure 2: The coefficients of the four complex eigenvectors of
(\ref{eq3}) and its complex conjugate. The four possibilities are 
$(r,z)\in\{(\xi_+,\tau_+),(\xi_-,\tau_-),(\xi_+^{-1},\tau_+),
(\xi_-^{-1},\tau_-)\}$.\bigskip

From this analysis one can derive a set of generators
for $\cH_{\PT}(\RR^2)$. From
linear combinations of (\ref{eq1}) (resp.\ (\ref{eq2}))
one obtains $B_1,B_2$ (resp.\ $B_3,B_4$) where
$$ B_1 = \sum_{i=0}^{19} \alpha_i(1), \quad B_2 = \sum_{i=20}^{39}
\alpha_i(1),\quad  
B_3 = \sum_{i=0}^{19} (-1)^i \alpha_i(1), \quad 
B_4 = \sum_{i=20}^{39}(-1)^i \alpha_i(1) .$$
The remaining $4$ generators are more complicated expressions 
but straightforward to derive from (\ref{eq3}). In particular, to
obtain two $2$-dimensional real irreducible representations of
$D_{10}$ one has to take the real and the imaginary part of (\ref{eq3}).

We mention that this method allows also to determine
a sub-group of finite index of the integer valued second cohomology. 
This amounts to
taking linear combinations of the above vectors so that the result
lies in the intersection of the above complement with $\ZZ^{40}$.
In that case one obtains integer irreducible representations of
$D_{10}$ which are $2$, $2$ and $4$ dimensional, cf.\ 
\cite{OrmesRadinSadun02}.    

\subsection{$k=1$}
A closed strongly $\PT$-equivariant $1$-form is
given by $f_1 dx_1+ f_2 dx_2$ where $f_i$ are smooth strongly
$\PT$-equivariant functions satisfying $\frac{\partial f_2}{\partial x_1} =
\frac{\partial f_1}{\partial x_2}$. A similar
construction as the one presented for $k=2$, namely one 
where one starts with a compactly supported
closed $1$-form and places it at all points of a uniformly
discrete set which is locally derived from $\PT$, 
produces only exact forms. This follows again from the
Poincar\'e Lemma.

To construct closed but non-exact $1$-forms 
we use the well-known Ammann lines
on the Penrose tilings. Ammann lines are five families of parallel lines
which  are locally derivable from a Penrose tiling \cite{GrSh}. 
The distance between two consecutive parallel Ammann lines of one
family is $1$ or $\tau_+$ (in suitable
units) and the sequence of distances forms a
Fibonacci sequence (called musical sequence in \cite{GrSh}). 
Fix one family $\F_1$ of Ammann lines
and let $(x,y)$ be orthonormal coordinates of $\RR^2$ such that $y$ is parallel
and $x$ normal to the Ammann lines of $\F_1$. In particular, a line $N$
normal to the Ammann lines of $\F_1$ is parametrized by $x$ and cut
into a Fibonacci sequence by its intersection points with these Ammann
lines. Let $\P_N$ be the set of intersection points in $N\cong \RR$
and $f:\RR\to\RR$ be a smooth strongly $\P_N$-equivariant function.
Then $(x,y)\mapsto f(x)dx$ is a closed strongly $\PT$-equivariant $1$-form
over $\RR^2$. In the same way as we reasoned for the case $k=2$ above
one can construct two strongly $\P_N$-equivariant functions each one 
associated with a prototile in $N$, namely by placing a Dirac comb
on the centers of mass of either the unit intervals or
the intervals of length $\tau_+$. 
In the cohomology $\cH_{\P_N}^1(\RR)$ corresponding to the
  $1$-dimensional subsystem these two functions lead to two
  independent generators.
Considering all families of
Ammann lines this gives $10$ strongly $\PT$-equivariant $1$-forms.
However, one may show that in $\cH^1_{\PT}(\RR^2)$ only $4$ of them
are independent. In fact, the appearance of these $1$-forms is related
to the higher dimensional lattice used to construct the Penrose tiling
via the cut \& projection method and somewhat independent of the details of the
tiling which originate from the choice of windows \cite{KPneu}.
 
From \cite{AP,OrmesRadinSadun02} we know (in combination with
Theorem~\ref{thm1}) that $\cH_{\PT}^1(\RR^2)\cong\RR^5$ and the latter
splits into a sum of a
$4$-dimensional irreducible $C_{10}$ module with a $1$-dimensional irreducible
$C_{10}$-module. The $4$-dimensional irreducible $C_{10}$ module is
given by the above construction using the Ammann lines. The remaining
one is related to a locally derivable orientation of the edges of the
Penrose tiling \cite{Gaehlerpriv}, it will not be discussed here. 

\section{$\P$-equivariant functions and the continuous hull of
  $\P$}\label{sect4} 

For the definition of $\P$-equivariant cohomology we have considered
smooth real valued strongly $\P$-equivariant functions over $\RR^n$. 
To explain the relation with earlier definitions of cohomology groups
for $\P$, in particular that involving the continuous hull, 
we now consider continuous complex valued functions.
\begin{definition}
The algebra of complex continuous $\P$-equivariant functions
$C_\P(\RR^n)$ is the closure in the supremum norm of the space of
complex continuous strongly $\P$-equivariant functions over $\RR^n$ with
pointwise multiplication.
\end{definition}
$C_\P(\RR^n)$ is a commutative $C^*$-algebra with unit and
by the Gelfand-Naimark theorem there exists a 
compact space $X_\P$ such that $C_\P(\RR^n)$ is isomorphic to  $C(X_\P)$.
The simplest example is certainly $\P=\emptyset$ in which 
$C_\emptyset(\RR^n) \cong \CC$ and so $X_\emptyset$ is a single point.
Another simple example but with $\P$ being a Delone set is a regular lattice
$\Gamma\subset \RR^n$, then $C_\Gamma(\RR^n) \cong C(\TT^n)$, i.e.\
$X_\Gamma=\TT^n=\RR^n/\Gamma$, the $d$-torus. 
In general, $X_\P$ is a lot more complicated.

The set of subsets of $\RR^n$ which have \flc\ carries a metric
$$D(S,S') = \inf \{\frac{1}{r+1}|\exists x,x'\in\RR^n, |x|,|x'|\leq
\frac{1}{r}: B_r\cap (S-x)=B_r\cap (S'-x')\}.$$
The completion of the set of translates of $\P$ w.r.t.\ this metric,
$$\M_\P:=\overline{\{\P-x|x\in\RR^n\}}^D,$$
is the continuous hull of $\P$, see e.g.\ \cite{AP}. 
By construction $D(S,S-x)$ is of order $|x|$ if the norm $|x|$ of $x$
is small. In particular, $\M_\P$ is connected. If $\P$ is a Delone set
then all elements of $\M_\P$ can be interpreted as Delone sets. (If
$\P$ is not relatively dense then $\M_\P$ contains the empty set as element.)

\begin{lemma}\label{lem1} Let $\P$ be a Delone subset of $\RR^n$ which has
\flc. Then $X_\P\cong \M_\P$. More precisely, $\sigma:C(\M_\P)\to
  C_\P(\RR^n)$,
$$\sigma(f)(x) =  f(\P-x)$$
is an algebra isomorphism.
\end{lemma}
\bew\ The space $\M_\P$ is in general not a manifold but a foliated space in
the sense of \cite{MooreSchochet88}. 
Its leaves are the sets of translates of points. 
They are locally homeomorphic to $\RR^n$. Therefore,
$\sigma(f)$ is continuous. 

In the transverse direction $\M_\P$ is totally
disconnected. In the (completely) periodic cases the
transversals consist of finitely many points. 
For (completely) aperiodic Delone sets the transversals are Cantor sets. If
we consider the canonical transversal
$\Omega_\P=\{\omega\in\M_\P|0\in\P\}$ we can describe its topology
as that being generated by the clopen sets $U_{P,p}$ where $p\in P\subset\P$,
$P$ a finite subset and $U_{P,p}=\{\omega\in\Omega_\P|P-p\subset \omega\}$.    
The topology of $\M_\P$ is then generated by
$U_{P,p,\epsilon,y}=\{\omega\in\M_\P| \exists
x\in B_\epsilon(y):P-p-x\in\omega\}$ (cf.\ \cite{FHKphys}). 
Moreover, we may restrict
$\epsilon>0$ to values smaller than the
minimal distance ${r_0}$ of two points in $\P$. For such $\epsilon$,
$U_{P,p,\epsilon,y}$ is homeomorphic to $U_{P,p}\times
B_\epsilon(y)$. It follows that
$C(\M_\P)$ is generated by continuous
functions which are supported on sets 
$U_{P,p,\epsilon,y}$, $\epsilon<r_0$. 
Such functions are of the form
$f_{P,p,\rho}=\delta_{P,p}*\rho$ where $\delta_{P,p}$ 
is a Dirac comb placed on
the set $\{x\in\RR^n|P-p\in\P-x\}$ and $\rho$ is continuous and 
has support inside $B_{r_0}(y)$. But $\sigma(f_{P,p,\rho})$ is strongly
$\P$-equivariant, the value for $r$ in Definition~\ref{def1} 
being at most $r_0+|y|$.
This shows that $\sigma$ maps $C(\M_\P)$ into $C_\P(\RR^n)$. From denseness
of the leaf through $\P$ follows that $\sigma$ is injective.

If $g:\RR^n\to\CC$ is a strongly
$\P$-equivariant continuous function define
$\tilde g:\M_\P\to \CC$ by $\tilde g(\omega) = g(\P-x)$ where $x$ is such
that $B_r\cap (\P-x)=B_r\cap\omega$ (the value for $r$ from 
Definition~\ref{def1}). Then $\tilde g$ is continuous and
$\sigma(\tilde g) = g$. 
\eb\bs

Let us indicate how the last lemma helps to prove Theorem~\ref{thm1}, 
i.e.\  that the \ch ech cohomology of $\M_\P$ with coefficients in $\RR$ is 
isomorphic to
$\spH(\RR^n)$. Although $\M_\P$ may not be a manifold its leaves are
and one can define functions which are smooth
in the direction tangential to the leaves.
This leads to the definition of 
tangential differential forms and consequently of tangential
cohomology \cite{MooreSchochet88}.
The inverse of the map $\sigma$ above identifies the complex of strongly
$\P$-equivariant differential forms with a sub-complex of the complex of
tangential differential forms. $\P$-equivariant cohomology equates
therefore with the cohomology of a sub-complex of tangential forms.
On the other hand there exists an analogue of the  
\ch ech-de Rham complex \cite{BottTu82} providing us with a homomorphism from 
the \ch ech cohomology of $\M_\P$ (with real coefficents) to 
tangential cohomology of $\M_\P$. Its image is precisely
$\P$-equivariant cohomology.
\bigskip

\noindent
{\bf Acknowledgements.}
I wish to thank Ian Putnam for discussions which led to the ideas underlying
this article.


\begin{thebibliography}{FHK02}

\bibitem[AP98]{AP}
J.E. Anderson and I.F. Putnam, \emph{Topological invariants for substitution
  tilings and their associated {$C^*$}-algebras}, Ergod. Th. and Dynam. Sys.
  \textbf{18} (1998), 509--537.

\bibitem[BSJ91]{BSJ}
M.~Baake, M.~Schlottmann, and P.D. Jarvis, \emph{Quasiperiodic tilings with
  tenfold symmetry and equivalence with respect to local derivability}, J.
  Phys. A \textbf{24} (1991), 4637--4654.

\bibitem[BT82]{BottTu82}
R.~Bott and L.W. Tu, \emph{Differential forms in algebraic topology},
  Springer-Verlag, New York, 1982.

\bibitem[Con90]{Co}
A.~Connes, \emph{{G\'eom\'etrie Non Commutative}}, InterEditions (Paris), 1990.

\bibitem[Con94]{Cone}
A.~Connes, \emph{{Non Commutative Geometry}}, Academic Press, 1994.

\bibitem[FH99]{FoHu}
A.H. Forrest and J.~Hunton, \emph{The cohomology and {K}-theory of commuting
  homeomorphisms of the {C}antor set}, Ergod. Th. and Dynam. Sys. \textbf{19}
  (1999), 611--625.

\bibitem[FHK02]{FHKphys}
A.H. Forrest, J.~Hunton, and J.~Kellendonk, \emph{Cohomology of canonical
  projection tilings}, Comm. Math. Phys. \textbf{226} (2002), no.~2, 289--322.

\bibitem[G{\"a}h]{Gaehlerpriv}
F.~G{\"a}hler, privat communication.

\bibitem[GS87]{GrSh}
B.~Gr\"unbaum and G.C. Shephard, \emph{Tilings and patterns}, Freeman and
  Company (New York), 1987.

\bibitem[Kel95]{Ke2}
J.~Kellendonk, \emph{Non commutative geometry of tilings and gap labelling},
  Rev. Math. Phys. \textbf{7} (1995), 1133--1180.

\bibitem[Kel97]{Ke5}
J.~Kellendonk, \emph{The local structure of tilings and their integer group of
  coinvariants}, Commun. Math. Phys. \textbf{187} (1997), no.~1, 115--157.

\bibitem[KP]{KPneu}
J.~Kellendonk and I.F. Putnam, in preparation.

\bibitem[MS88]{MooreSchochet88}
C.C. Moore and C.~Schochet, \emph{Global analysis on foliated spaces},
  Mathematical Sciences Research Institute Publications, 9, Springer-Verlag,
  New York, 1988.

\bibitem[ORS02]{OrmesRadinSadun02}
N.~Ormes, C.~Radin, and L.~Sadun, \emph{A homeomorphism invariant for
  substitution tiling spaces}, Geom. Dedicata \textbf{90} (2002), 153--182.

\end{thebibliography}
\providecommand{\bysame}{\leavevmode\hbox to3em{\hrulefill}\thinspace}

\end{document}